\documentclass[letterpaper,12pt]{jpconf}
\usepackage[utf8]{inputenc}
\usepackage{graphicx}
\usepackage{amsfonts}
\graphicspath{{Figures/}}
\usepackage[breaklinks=true,colorlinks=true,linkcolor=blue,urlcolor=blue,citecolor=blue]{hyperref}
\usepackage{soul}
\usepackage{cite}
\usepackage{float}
\usepackage{multirow}
\begin{document}

\title{Molecular dynamics simulation of behaviour of water in nano-confined ionic liquid--water mixtures}

\author{B Docampo-\'{A}lvarez$^{1}$, V G\'{o}mez-Gonz\'{a}lez$^{1}$, H Montes-Campos$^{1}$, J~M~Otero-Mato$^{1}$, T M\'{e}ndez-Morales$^{1}$, O Cabeza$^{2}$, L J Gallego$^{1}$, R~M~Lynden-Bell$^{3}$, V~B~Ivani\v{s}t\v{s}ev$^{4}$, M V Fedorov$^{5}$ and L M Varela$^{1}$}
\address{$^1$Departamento de F\'{i}sica da Materia Condensada, Facultade de F\'isica, Universidade de Santiago de Compostela, Campus Vida s/n, E-15782 Santiago de Compostela, Spain}

\address{$^2$Departamento de Física, Facultade de Ciencias, Universidade da Coru\~{n}a, Campus A Zapateira s/n, E-15071 A Coru\~{n}a, Spain}

\address{$^3$Department of Chemistry, University of Cambridge, Lensfield Road, CB2 1EW Cambridge, UK}

\address{$^4$Institute of Chemistry, University of Tartu, Ravila 14a, 50411 Tartu, Estonia}

\address{$^5$Department of Physics, Scottish Universities Physics Alliance (SUPA), Strathclyde University, John Anderson Building, 107 Rottenrow East, G4 0NG Glasgow, UK}

\eads{\mailto{luismiguel.varela@usc.es}}

\begin{abstract}

This work describes the behaviour of water molecules in 1-butyl-3-methylimidazolium tetrafluoroborate ionic liquid under nanoconfinement between graphene sheets.  By means of molecular dynamics simulations, an adsorption of water molecules at the graphene surface is studied. A depletion of water molecules in the vicinity of the neutral and negatively charged graphene surfaces and their adsorption at the positively charged surface are observed in line with the preferential hydration of the ionic liquid anions.   The findings are appropriately described using a two-level statistical model.  The confinement effect on the structure and dynamics of the mixtures is thoroughly analyzed using the density and the potential of mean force profiles, as well as by the vibrational densities of states of water molecules near the graphene surface.  The orientation of water molecules and the water-induced structural transitions in the layer closest to the graphene surface are also discussed.  
\end{abstract}

\noindent{\it Keywords}: Ionic liquids, electric double layer, molecular dynamics simulation, graphene, interfacial layer structure.


\section{Introduction}

The recent search for safe and ``green'' energy storage devices has opened the door to designing systems using ionic liquids (ILs) as novel electrolytes \cite{Vatamanu2013ude,Fedorov2014uyy,Macfarlane2014ugy}. Their unique properties \cite{Wasserscheid2003wep,Ohno2005ujg} allow them to be considered as the solvents of the future in many fields \cite{Plechkova2008tnl,Macfarlane2007tux}.  On the other hand, the carbon-based nanostructures are also seen as very promising materials for energy storage. Not only due to their high accessibility and a low preparation cost, but also due to their high surface area, an easy processability and controllable pore size \cite{Zhai2011vkf,Zhang2009uay,Simon2013ucv,Beguin2014tsc}.  Thus, the design of the advanced electrochemical and electromechanical technologies is expected to go through the research of carbon-based materials and the novel amphiphilically nanostructured green designer solvents.  

In device applications ILs are frequently incorporated in polymer or confined porous matrices, producing ``ionogels'' \cite{Singh2014tpt}.  In electrochemical double layer capacitors, actuators and sensors, these electrolytes were found to exhibit a more efficient behaviour under some confinement conditions \cite{Merlet2012wne}. Partly, it is because of the carbon electrodes porosity that leads to a larger surface area.  The nanoconfinement of liquids, specifically ILs, affects dramatically their physical properties when compared to bulk \cite{Perkin2012tov}.  The optimisation of these applications of ILs requires a fundamental knowledge of these dense ionic solvents in a nanoconfined environment. Although a great interest on nanoconfined ILs has already arisen, the study of the confinement of ILs is still in its infancy, according to the Singh \textit{et al.} review of the science and technological applications of ILs confined in nanopores \cite{Singh2014tpt}.  Perkin, who has recently analyzed the studies on structure, dynamics, and colloidal forces in confined ILs \cite{Perkin2012tov}, finds the effect of confining surfaces on the structure of ILs or the changes in the ordering of cations and anions when charging the electrodes to be an opened question.  

Computer simulations play a significant role in the description of these systems.  Merlet \textit{et al.} reported molecular dynamics (MD) simulations on the microscopic mechanism of the capacity enhancement of IL-based superconductors and the structure of an IL adsorbed inside realistic microporous carbon electrodes \cite{Merlet2012wne}.  The structure and dynamics of an imidazolium-based IL inside a slit graphitic nanopore and realistic activated carbon were also contributed by Rajput \textit{et al.} \cite{Rajput2014vnw}.  A slower dynamics was found for the ions close to the surfaces, and the authors found that the structure and dynamics of confined ILs inside porous materials with the heterogeneity of pore size, shape and inter-connectivity show significant departure from the properties of ILs confined inside the simpler slit nanopores. For the pores of simple geometries, like the slit nanopore, the interplay between the potential-dependent layering of IL ions (see Ref. \cite{Ivanistsev2014tqg}) and the pore width has a pronounced effect on the interfacial capacitance \cite{Feng2013vwj}. Herewith, in experiments the slit-pore model is often assumed to be the most suitable to describe the pores in micro/mesoporous carbon materials \cite{Kurig2016wts}. Nanoconfinement in an ideal slit nanopore under applied potential can be investigated by using the recently developed graphene surface force balance method \cite{Britton2014twk}.  This experimental technique allows elucidating the effect of an added Li$^+$ on the interfacial structure of an IL with molecular resolution.  In line with the experimental results, M\'endez-Morales \textit{et al.} \cite{Mendez-morales2014uev} and Ivani\v{s}t\v{s}ev \textit{et al.} \cite{Ivanistsev2016vrv} have performed MD simulations of the electric double layer in mixtures of ILs and lithium and potassium salts of electrochemical interest, proving the existence of high energy barriers for Li$^+$/K$^+$ transport to electrodes in nanoscopic cages. Thus, it is possible to investigate the nanoconfined IL--solute mixtures with complementary computational and experimental methods, and then extrapolate the results to porous carbon materials.

It is well-known that the presence of water in ILs is almost inevitable due to their highly hygroscopic character \cite{Chen2013tnc,Cuadrado-prado2009wqq}. This fact can have a profound impact on their properties and the applications of these solvents.  The level of water uptake from the atmosphere, which is usually governed by the nature of the anion, was found to affect several electrochemically relevant properties such as viscosity \cite{Widegren2005wre} and electrical conductivity \cite{Rilo2009ufw,Rilo2013tvp}.  Moreover, Compton and co-workers \cite{Omahony2008tmf} found a decrease of the electrochemical windows in twelve different ILs with an increasing water content, which is of fundamental importance when it comes to selecting some particular IL-based electrolytes for electrochemical applications.  

The existence of water in confined geometries was extensively studied in several fields of chemistry, biology, materials science, and technological applications \cite{Fayer2010tsk, Zhou2012vmt}. However, despite the aforementioned unavoidable presence of water in ILs, the nanoconfined IL--water mixtures were probed only in a few computational and experimental studies \cite{Espinosa-marzal2014vot,Feng2014vrd}.  
The very first computational study of water in ILs at an electrified interface was performed by Kornyshev and co-workers\cite{Feng2014vrd}.  The authors investigated the adsorption of water molecules on planar graphite surface from the water mixtures with imidazolium-based ILs by using MD simulations.  
The results revealed that the water accumulation at the interface is related to the water molecule's propensity to follow the position of the maximal absolute value of the electrical field, the association with the surrounding ions, and the tendency to occupy spaces of a lower local density.  Kornyshev and co-workers demonstrated that the balance between these factors under varying surface charge density can greatly affect the microscopic environment of the water molecules near the electrode surface, and thus concluded that in the real system the energetics of water adsorption can be altered by the electrode potential.

Undoubtedly, the presence of water molecules at the interface can have crucial effects on the use of ILs as electrolytes in electrochemical and electromechanical devices, \textit{e.g.} double layer capacitors and actuators.  Besides the applied potential, the characteristics of the electrode surface, such as porosity and roughness, are known to affect the adsorption.  In the work described here we extend the investigation performed by Kornyshev and co-workers by exploring the effects of confinement on the water adsorption from IL--water mixtures.  To the best of our knowledge, the effect of nanoconfinement on the structure and dynamics of IL--water mixtures has not been previously reported.  

The structure of the paper is as follows. The essential technical details of the computational methods used are sketched in Section 2, the results are presented and discussed in Section 3, and in section 4 we summarise our key findings. 


\section{Computational methods}

\begin{figure}[ht]
  \begin{center}
  \includegraphics[width=3.25in]{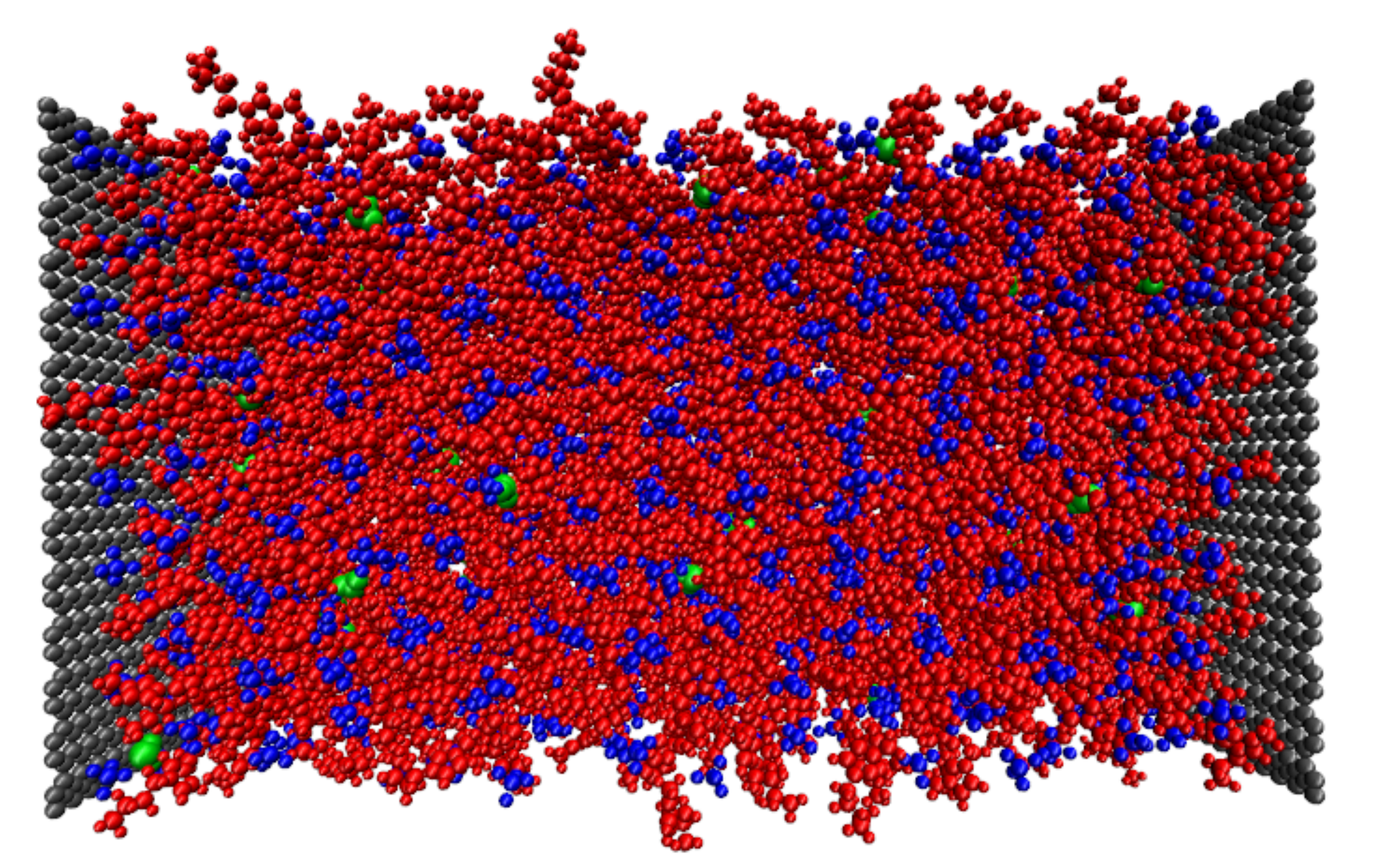}
  \end{center}
  \caption{Snapshot showing the 10.04 nm simulation box with charged walls; red: cation atoms; blue: anion atoms; green: water molecules. The latter are slightly enlarged for clarity purposes.}  
\label{fig:figure_box}
\end{figure}

MD simulations were carried out using the Gromacs 5.0 software \cite{Hess2008vel,Pronk2013vhm,Van_der_spoel2010wfn,Lindahl2001tps,Berendsen1995try} in an $NVT$ ensemble with a temperature of 298~K maintained with a V-rescale thermostat \cite{Bussi2007vdd}.  To avoid the system from being ``trapped'' in a local minimum due to a very high viscosity of the IL, we increased the temperature of the mixture up to 600~K by using simulated annealing prior to the production run.   The systems studied consisted of [BMIm][BF$_4$] (1-butyl-3-methylimidazolium tetrafluoroborate) with 5\% molar percentage of water (equivalent to about 4200 ppm of water in weight) and two rigid graphene sheets in a simulation box with periodic boundary conditions along the limits of the graphene sheets. The IL was parameterized with the OPLS-AA force field \cite{Spoel2005wyt}, which includes intramolecular terms for bond stretching, angle bending and dihedral torsion, as well as van der Waals and Coulomb forces.  For water molecules, we employed the TIP5P model of Mahoney and Jorgensen \cite{Mahoney2000tfp}, which, in addition to the Lennard-Jones (LJ) centre placed on the oxygen and the charges located at the hydrogen atoms, places two partial charges representing the lone pairs of the oxygen atom.  Finally, rigid graphene sheets of varying sizes, adjusted to accommodate all molecules while allowing for periodic boundary conditions with the desired cell size, were created with the help of the Visual Molecular Dynamics (VMD) package \cite{Humphrey1996tsu}. One sheet was placed parallel to the $x,y$-plane at $z=0$ and the others at $z=L$, with $L$ values of 1.70, 2.62, 3.06, 4.18, 5.77, 8.12, 10.04 and 11.42~nm. The LJ parameters used for the carbon atoms were $\sigma_\mathrm{C}=3.55\times 10^{-1}$~nm and $\epsilon_\mathrm{C}=2.9288 \times 10^{-1}\,\mathrm{kJ\,mol}^{-1}$. Each simulation had 950 ion pairs and 50 water molecules, in order to have enough water molecules to provide meaningful statistical data.  In the case of the 11.42~nm simulation, the amounts were increased to 70 water molecules and 1330 ion pairs to keep the graphene sheet, and therefore the simulation cell, large enough. Additionally, a second 11.42~nm-long system simulation box without any water molecules was prepared and simulated for comparison. The precise number of particles of each species in the system are included in Table \ref{label}.
\begin{table}
\caption{\label{label}Simulation box width (in nm) and number of IL pairs, water molecules, and electrode C atoms.}
\begin{center}
\item[]\begin{tabular}{@{}llll}
\br
Box width & IL pairs & Water & C atoms \\
\mr
1.7 & 950 & 50 & 8260 \\
2.62 & 950 & 50 & 4968 \\
3.06 & 950 & 50 & 4200 \\
4.18 & 950 & 50 & 3024 \\
5.77 & 950 & 50 & 2160 \\
8.12 & 950 & 50 & 1560 \\
10.04 & 950 & 50 & 1288 \\
11.42 & 1330 & 70 & 1500 \\
\br
\end{tabular}
\end{center}
\end{table}
Moreover, a scheme with simulation box is included in Fig. \ref{fig:figure_box}.

Ions were packed into the simulation box with Packmol \cite{Martinez2009uoq} using the density values obtained in $NPT$ simulations of bulk [BMIm][BF$_4$]--water mixture. Thereafter each system was annealed at 600~K ($dt = 2\,\mathrm{fs}$ for $1\,\mathrm{ns}$), equilibrated ($dt = 2\,\mathrm{fs}$ for $10\,\mathrm{ns}$), and simulated ($dt = 2\,\mathrm{fs}$ for $5\,\mathrm{ns}$).  Long-range electrostatic interactions were treated by the Particle Mesh Ewald method \cite{Frenkel2002wwr}. In the runs with charged surfaces, the charges on the carbon sites on the graphene sheet were chosen to give surface charge densities of $\pm 1$~e/nm$^2 = \pm16$~$\mathrm{\mu}$C/cm$^2$.  The left-hand and right-hand surfaces carried equal and opposite charges.  The results were analyzed using a combination of Gromacs' tools and custom scripts employing the MDAnalysis Python package \cite{Michaud-agrawal2011wef}.


\section{Results and discussion}

\subsection{Density distribution in the nanoconfined IL--water mixture}


\begin{figure}[ht]
  \begin{center}
  \includegraphics[width=3.25in]{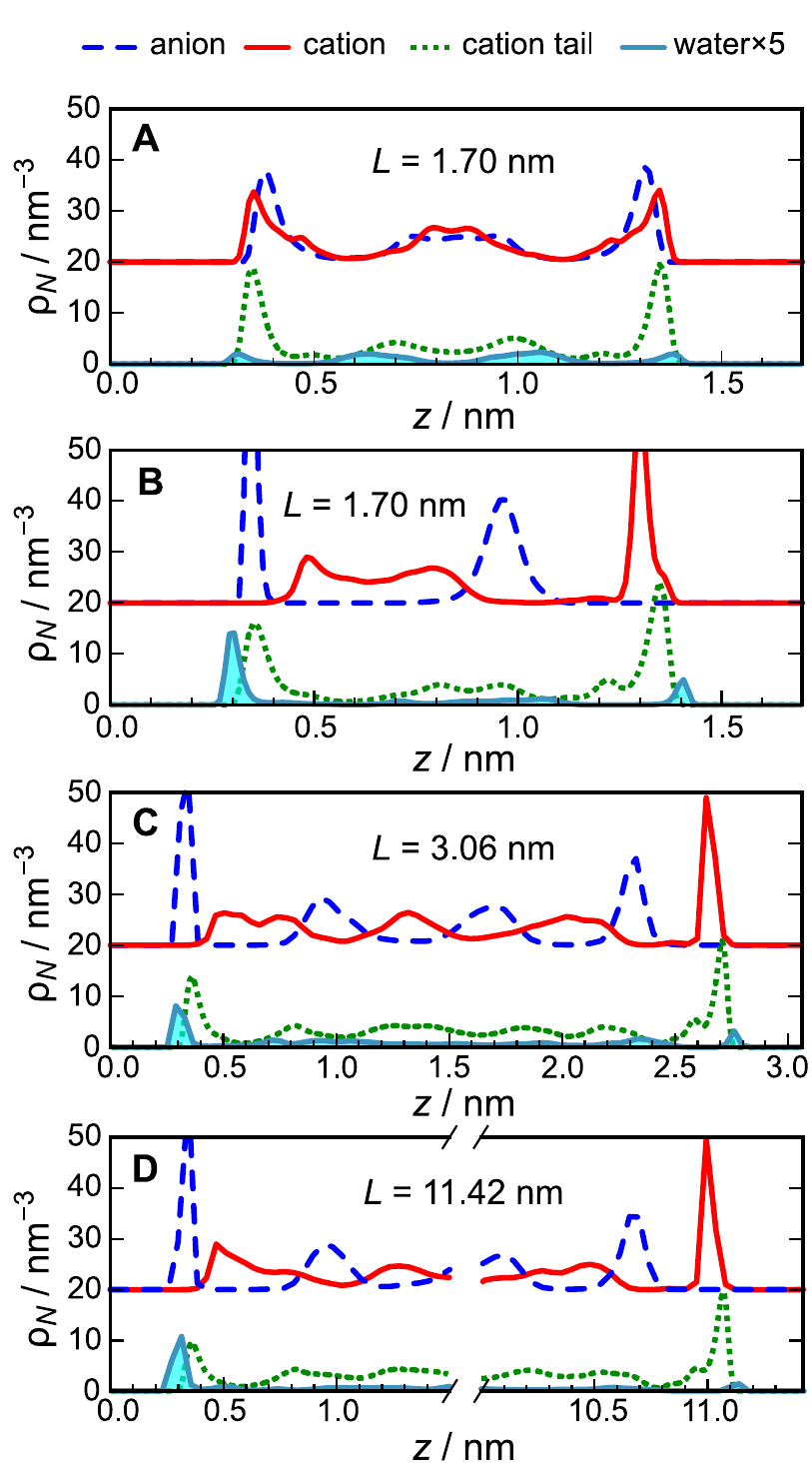}
  \end{center}
  \caption{Density profiles for anion (offset by 20~nm$^{-3}$), cation (offset by 20~nm$^{-3}$), cation tail and water molecule (solid curve, magnified by 5) in systems with different distance between the graphene sheets ($L$).  (A) system with neutral graphene sheet situated as $z = 0$ and $z = L = 1.70\,\mathrm{nm}$; (B)--(D) systems with a positively charged graphene sheet situated at $z = 0\,\mathrm{nm}$ and a negatively charged graphene sheet situated at $z = L$.}
  \label{fig:figure_density}
\end{figure}

Number densities of the different species present in the mixture along the $z$-axis perpendicular to the graphene sheets were calculated for all simulation boxes and a selection of the results are shown in Fig.  \ref{fig:figure_density}.  In the absence of charge on the surfaces or significant nanoconfinement, the surface interactions affect the anion and cation curves only up to a distance of approximately 1.5--2~nm, recovering bulk liquid ordering beyond that distance \cite{Mendez-morales2014uev}.  This interfacial structure consists of a 0.2~nm gap (corresponding to the distance of closest approach of the graphene carbon sites used to define the densities) followed by close, almost overlapping peaks of high local density (3 to 5 times the values of bulk density of $3\,\mathrm{nm}^{-3}$) near the surface for each of the ionic species.  A low-density zone and a second lower peak follow, with a third peak being still distinguishable from the bulk.  This layering is reinforced for confinement below 4~nm, where something similar to a bulk IL is not found, and the regions directly affected by the presence of the graphene sheets overlap.  Water seems to accumulate in the low-density regions, as it can be seen in Figure \ref{fig:figure_density}A, resulting in less water near the graphene sheets.

On the other hand, in simulations with charged surfaces a strongly heterogeneous structure of alternating ionic layers appeared, which is common for electrified IL--electrode interfaces \cite{Li2014ulb,Ivanistsev2014wyn}.  Figures \ref{fig:figure_density}B--D show the evolution of this structure with the increase $L$, \textit{i.e.} the distance between the graphene sheets.  As can be seen, the characteristic ionic density peaks near the surfaces are preserved in all of the cases studied.  Note that water shows a marked preference for the positively charged surface, with only a residual presence near the negatively charged surface.  This is probably associated with the stronger interaction of water molecules with the IL anions than with the cations, so water molecules tend to be accommodated close to the anions.  This seems to be the case for all studied systems with the variable distance between the graphene sheets.  However, water is also observed close to the negatively charged surface, 
where hydrogen atoms of the water interact with the graphene sheet, and the water dipole points away from the surface (see section \ref{sec:3.1}) Indeed, MD simulations of sheared water molecules between graphene sheets \cite{Ye2012woj} revealed that charging the graphene surface leads to enhanced interactions  between the carbon atoms and the water molecules.  In the neutral case, water is also present near the graphene sheet, but this is probably due to the presence of anions in the first adsorbed layers.


\subsection{Energetics of water adsorption}


\begin{figure}
  \begin{center}
  \includegraphics[width=3.25in]{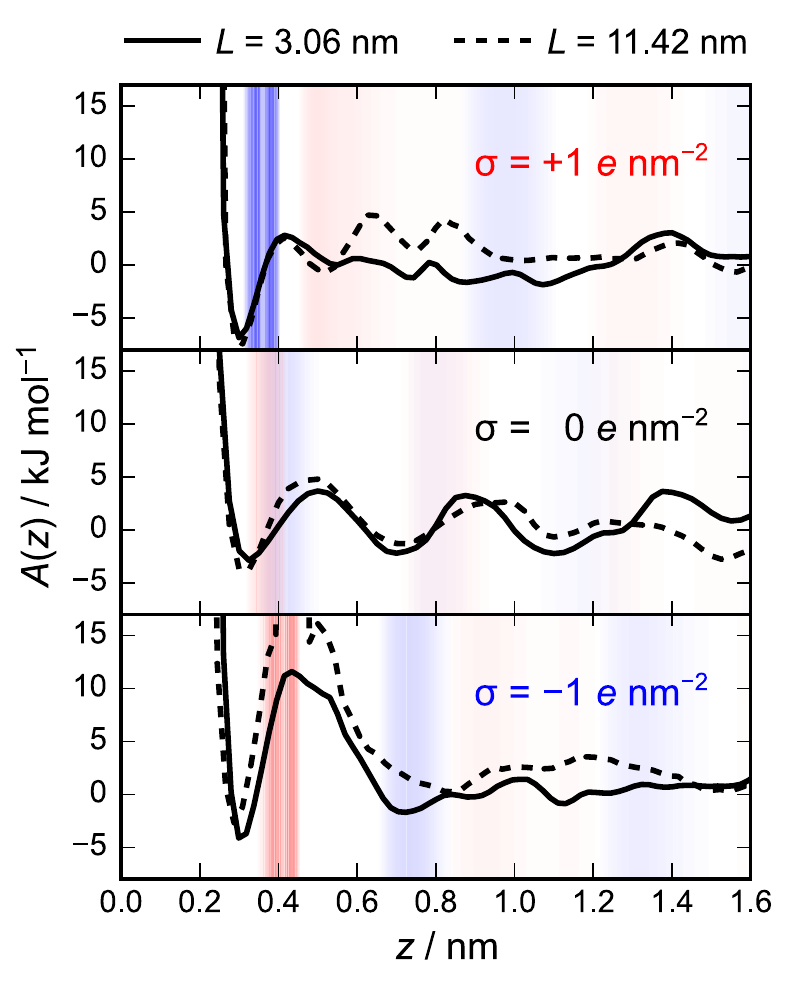}
  \end{center}
  \caption{PMF profiles of water molecules as a function of distance from the positively, neutral and negatively charged surfaces, for confined ($L = 3.06$~nm) and non-confined ($L = 11.42$~nm) cases.  The background represent the number density profiles for anion and cation represented as gradients of semi-transparent blue (darker) and red (lighter) colours, respectively.}
  \label{fig:figure_pmf}
\end{figure}

Figure \ref{fig:figure_pmf} shows the potential of the mean force (PMF) profiles for water, calculated for all $L$ from the above discussed number density curves as
\begin{equation}
A(z)=-k_\mathrm{B}T\ln\frac{\rho(z)}{\rho_\mathrm{bulk}}\,,
\end{equation}
where $k_\mathrm{B}$ is Boltzmann's constant and $T$ absolute temperature.  
In all cases, we observe a minimum of the PMF close to the graphene sheet, wherein this contact minimum is deeper by ~5 kJ mol$^{-1}$ while the free energy barrier is lower for the positively charged surface than for the neutral or negatively charged surfaces.  The effect of confinement is also seen in Fig.  \ref{fig:figure_pmf}.  The probability of water being adsorbed at the positively charged surface is increased for narrower $L$ while this is not the case for the neutral and negatively charged surfaces.  In the latter case, confinement leads to a significant increase of the barrier for water molecules to be transferred to the closest layer.  The density of anions is obviously larger close to the positively charged surface so that a much closer packing of IL moieties is expected in this layer than at the negatively charged surface.  This is probably behind the larger number of minima observed in $A(z)$ in that region.  

The average energies of anions and cations ($\langle E_\mathrm{ion}\rangle$, Table \ref{table:PMF}) for the interfacial regions were obtained as the mean value of the PMF in the first layer, \textit{i.e.} in the region comprised between the surface plane and the first maximum of $A(z)$.  
Except for the case of the negatively charged surface, $\langle E_\mathrm{anion}\rangle$ decreases once water is added.  In the case of the cation, the average energy even increases for the neutral surface.

\begin{table}[ht]
\caption{\label{table:PMF}Average energies for anions and cations ($\langle E_\mathrm{ion}\rangle$) in the first surface layer along the $z$-axis for the non-confined ($L = 11.42$~nm) pure IL and 5$\%$ IL--water mixture.}
\begin{center}
\item[]\begin{tabular}{lllll}
\br
 & & \begin{tabular}[x]{@{}c@{}}$\sigma$\\ $e$~nm$^{-2}$\end{tabular} & \begin{tabular}[x]{@{}c@{}}$\langle E_\mathrm{anion}\rangle$\\ kJ mol$^{-1}$\end{tabular} & \begin{tabular}[x]{@{}c@{}}$\langle E_\mathrm{cation}\rangle$\\ kJ mol$^{-1}$\end{tabular} \\ 
\mr
\multicolumn{2}{r}{\multirow{3}{*}{$[\mathrm{BMIM}][\mathrm{BF}_4]$}} & $-1$ & $+1.8$ & $-4.5$ \\
\multicolumn{2}{l}{}                          & $\phantom{+}0$ & $-0.1$ & $\phantom{-}0.0$ \\
\multicolumn{2}{l}{}                          & $+1$ & $-1.5$ & $+0.4$ \\
\mr
\multicolumn{2}{r}{\multirow{3}{*}{\begin{tabular}[x]{@{}c@{}}$[\mathrm{BMIM}][\mathrm{BF}_4]$\\ $+\mathrm{H}_2\mathrm{O}$\end{tabular}}} & $-1$ & $+1.8$ & $-4.5$ \\
\multicolumn{2}{l}{}                          & $\phantom{+}0$ & $-0.3$ & $+0.4$\\
\multicolumn{2}{l}{}                          & $+1$ & $-2.2$ & $+0.3$ \\ 
\br
\end{tabular}
\end{center}
\end{table}

\begin{figure*}[ht]
  \begin{center}
    \includegraphics[width=6in]{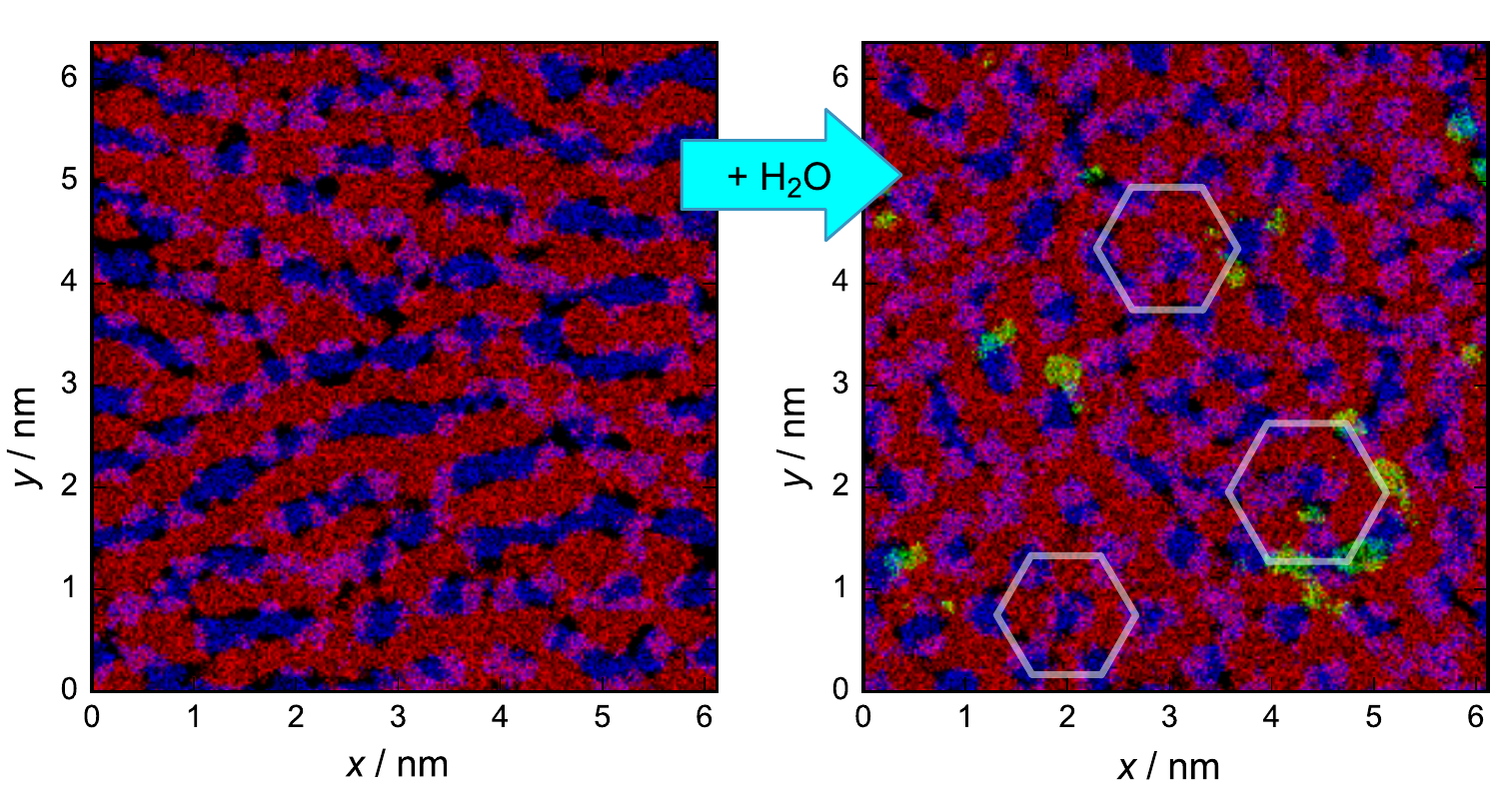}
  \end{center}
  \caption{Density map of anions (blue, darker) and cations (red, lighter) within 0.5~nm of a positively charged surface in absence of and with 5\% of water (green, light).  The overlaid hexagons highlight the change in the ions arrangement at the interface.}
  \label{fig:figure_densmap}
\end{figure*}

We have also analyzed the lateral $(x,y)$ structure of the first layer close to the positively charged surface of the non-confined ($L = 11.42$~nm) IL--water mixture.  We have observed that the lateral structure of the interfacial layer suffers a transition from ordered stripes to an ordered phase with hexagonal symmetry upon water addition (Fig.  \ref{fig:figure_densmap}). Recent progress in \textit{in situ} imaging and computer modelling of the potential-induced structural variations at IL interfaces has allowed to relate the transitions and the peaks at the differential capacitance vs. potential curves (see Ref. \cite{Rotenberg2015wqi} and references therein).  The differential capacitance alteration through dilution with organic solvents was theoretically interpreted and related to the ion concentration variation \cite{Bozym2015uyq}.  Similarly, accumulation of water molecules has direct impact on the ion concentration as well as on the interplay between ion packing constraints, Coulomb and van der Waals interaction within ILs, and the strength of interaction between IL--water mixture and surface.  A deeper insight into how the adsorption of water disrupts the polar--apolar networks of the IL may turn useful in designing novel IL--solvent mixtures for electrochemical and electromechanical devices with optimal power and energy density or mechanical properties.  A more detailed study of the features of this water-induced transition is beyond the scope of the present work and will be reported in due course.

\begin{figure}[ht]
  \begin{center}
  \includegraphics[width=3.25in]{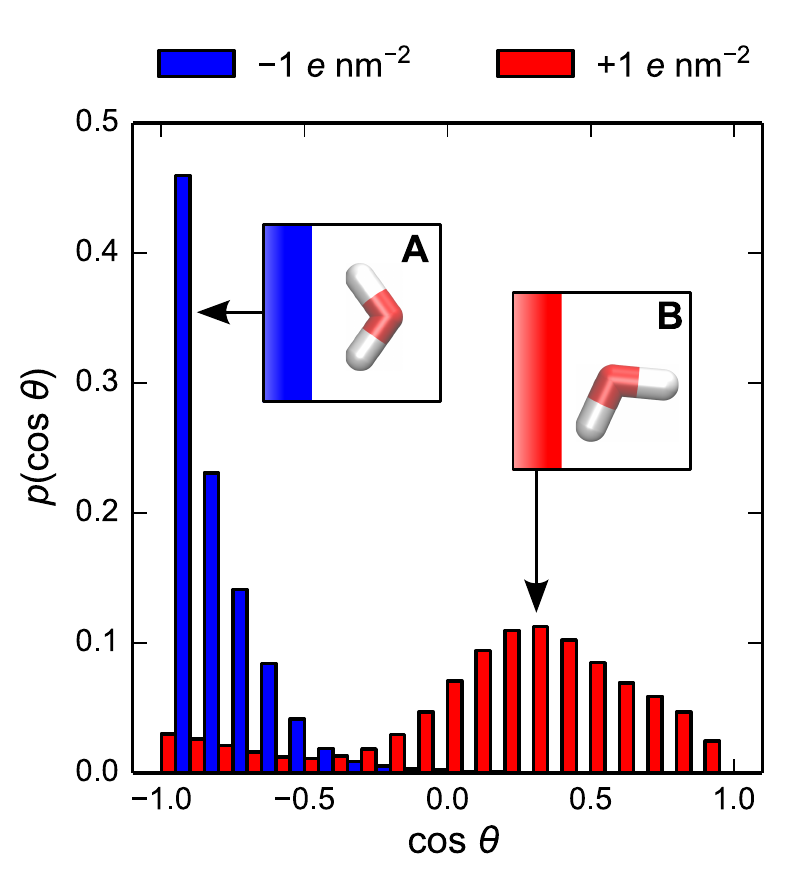}
  \end{center}
  \caption{Frequency distribution for $\cos \theta$ of water orientation in the first layer near charged surfaces for $L = 3.06$~nm.  The angle $\theta$ is taken as the one between the molecular dipole moment and the surface's normal vector.  Bars represent 0.1 wide intervals of cosine values.}
  \label{fig:figure_orientation}
\end{figure}

The translational and rotational states of water molecules in the layers close to the interface have also been analyzed.  The orientation of water molecules during simulations was recorded and histogrammed in Figure \ref{fig:figure_orientation}.  Results show that the orientation is essentially random in the bulk and tilted at the interface.  At the negatively charged surface, water is oriented with its dipole moment towards the surface, while at the positively charged surface it tilts towards nearby anions, consistent with the water locations highlighted in Figure \ref{fig:figure_densmap}.

\begin{figure}[ht]
  \begin{center}
  \includegraphics[width=3.25in]{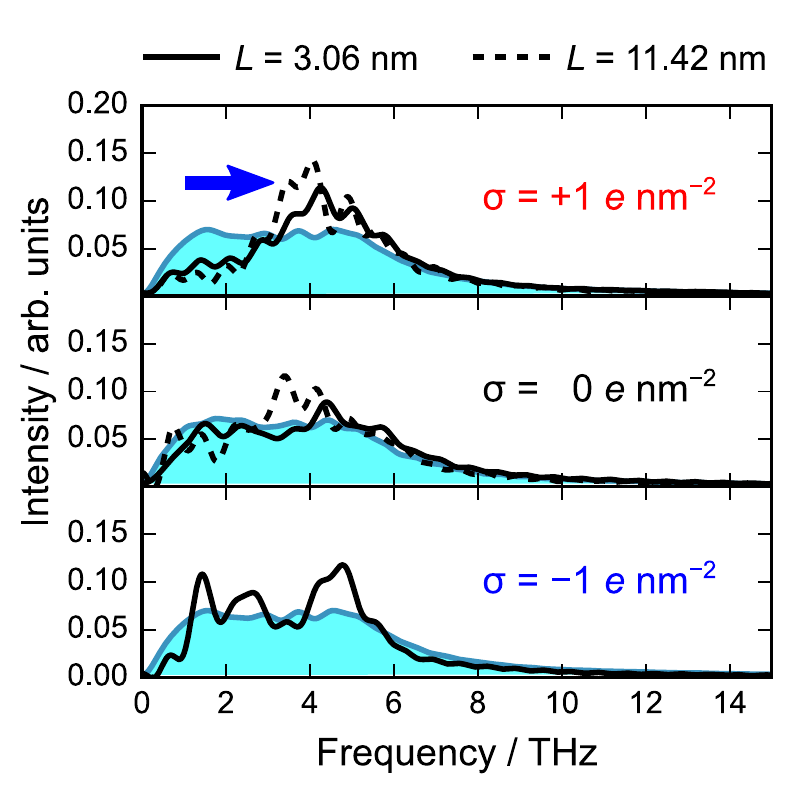}
  \end{center}
  \caption{Vibrational density of states for water molecules near the graphene sheets, with solid lines for the confined ($L = 3.06$~nm) and dashed lines for the non-confined ($L = 11.42$~nm) IL--water mixtures.  Lines with filled area represent the data for water molecules in the non-confined bulk.  The arrow indicates the blue shift in the vibrational density of states. }
  \label{fig:figure_vdos}
\end{figure}

The vibrational densities of states
\begin{eqnarray}
S(\omega)&=&\frac{1}{2\pi}\left| \int_{0}^{+\infty}dt e^{i\omega t}C(t)\right|^2\nonumber\\
&=&\frac{1}{2\pi}S(\nu)
\end{eqnarray}
of water molecules in the confined ($L = 3.06$~nm) and non-confined ($L = 11.42$~nm) IL--water mixtures were obtained from velocity autocorrelation functions
\begin{eqnarray}
C(t)=\frac{\left< \vec{v}(0) \cdot \vec{v}(t) \right>}{\left< \vec{v}(0) \cdot \vec{v}(0) \right>},
\end{eqnarray}
and the results are plotted in Figure \ref{fig:figure_vdos}.  There, it can be observed that confinement produces a blue-shift of the vibrational spectrum with respect to their bulk profiles, which is compatible with water molecules being more strongly bound in their cages in the interfacial environment.  This effect is more marked for the positively charged surfaces, where a much sharper displacement towards higher frequencies takes place. It is most likely related to a more ``caged'' state and a switch of orientation towards adsorbed anions.

\subsection{Statistics of water adsorption}
\label{sec:3.1}

To discuss the effect of confinement on water adsorption, we must take into account that an increased amount of water near the graphene sheet as the distance $L$ between surfaces decreases does not necessarily imply preferential adsorption.  As we are, effectively, considering a slice of the box near the surface, for a case of uniformly distributed water the probability of a single molecule being found at a distance closer than a limit $\delta$ would be

\begin{equation}
P_\mathrm{surf.} = \frac{ \int_0^\delta \mathrm{e}^{-\beta E(z)}\mathrm{d}z} {\int_0^{L} \mathrm{e}^{-\beta E(z)}\mathrm{d}z} = \frac{\int_0^\delta p_{0} \mathrm{d}z}{\int_0^{L} p_{0} \mathrm{d}z} = \frac{\delta} {L}\,, \label{eq:unidist}
\end{equation}
which is, in essence, the fraction of the volume considered.  Taking a constant $\delta$, in the absence of any specific interaction we could expect the fraction of water molecules near the surfaces to increase as the distance between the graphene sheets ($L$) decreases, following an inversely proportional law, until the limit case $L=2\delta$, where we would be considering the whole mixture as being near the surfaces.  In order to find out if specific interactions  between the species in the mixture induce accumulation or depletion of water near the surfaces, this result must be taken as a reference.

A simple model for studying preferential adsorption at the graphene surface considers different energy levels for water molecules in the bulk ($E_\mathrm{bulk}$) and in the interfacial regions ($E_\mathrm{\pm}$) with the energy curve taking the shape of rectangular wells/barriers depending on the sign.  Under the assumption that $E_{\pm}$ does not vary substantially as we decrease $L$, this allows for parameterisation and prediction of water behaviour using only the distance between graphene sheets.

\begin{equation}
E(z) = \left\{
\begin{array}{ll}
E_+, & z<\delta	\\
E_\mathrm{bulk}, & \delta < z < L - \delta \\
E_-, & L - \delta < z < L
\end{array}
\right.
\end{equation}

According to this model, the probability of finding one water molecule within a layer of width $\delta$ around the graphene sheets would be

\begin{equation}
P_{\pm} = \frac{\delta}{\delta (1 + \mathrm{e}^{\beta (\Delta E_{\pm} - \Delta E_{\mp})})+ (L-2\delta) \mathrm{e}^{\beta \Delta E_{\pm}}}\,,
\label{eq:fraction}
\end{equation}
where $\Delta E_{\pm} =E_{\pm} - E_\mathrm{bulk}$.  The neutral case $E_\mathrm{+}=E_\mathrm{-}$ simplifies to

\begin{equation}
P_\mathrm{neutral} = \frac{\delta}{2\delta + (L-2\delta) \mathrm{e}^{\beta \Delta E}}\,.
\end{equation}
Finally, for $\Delta E = 0$ we recover the previous, uniformly distributed case in Eq.  \ref{eq:unidist}.

\begin{figure}[ht]
  \begin{center}
  \includegraphics[width=3.25in]{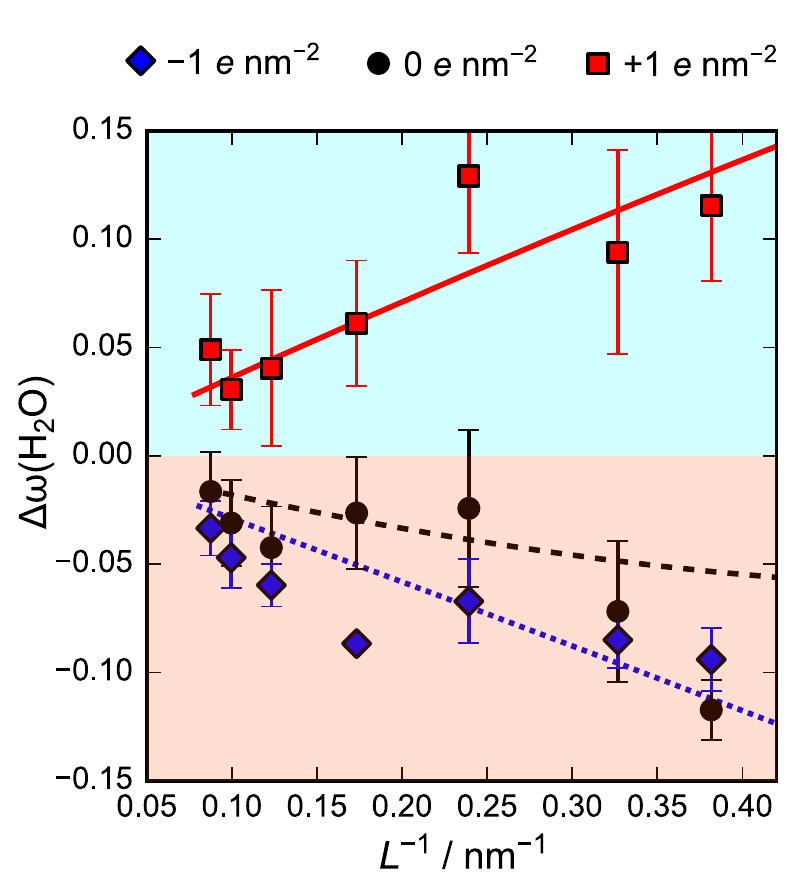}
  \end{center}
  \caption{Difference in the interfacial water fraction vs. distance between the graphene sheets ($L$).  Higher values indicate adsorption.  The lines show the respective fit to Eq.  \ref{eq:fraction}.}
  \label{fig:figure_fraction}
\end{figure}

\begin{table}[ht]
\caption{\label{table:energy}Fitted values for the depth of the potential energy well ($\Delta E$) near neutral and charged surfaces in Figure \ref{fig:figure_fraction}.  The mean integral value of the water PMF near each graphene sheet ($E_\mathrm{PMF}$) for $L=11.42$~nm is provided for comparison.}
\begin{center}
\item[]\begin{tabular}{@{}lll}
\br
\begin{tabular}[x]{@{}c@{}}$\sigma$\\ $e$~nm$^{-2}$\end{tabular} & \begin{tabular}[x]{@{}c@{}}$\Delta E$\\ kJ mol$^{-1}$\end{tabular}   & \begin{tabular}[x]{@{}c@{}}$E_\mathrm{PMF}$\\ kJ mol$^{-1}$\end{tabular}\\
\mr
$-1$ & $+0.85$ & $+0.94$ \\
$\phantom{+}0$ & $+0.48$ & $+0.39$\\
$+1$ & $-0.55$ & $-0.67$\\
\br
\end{tabular}
\end{center}
\end{table}

The number of water molecules within a distance $\delta=0.5$ nm from each surface was counted and averaged.  The chosen distance corresponds to the first PMF maximum for water, and, therefore, contains the whole first layer.  Figure \ref{fig:figure_fraction} shows, as an inverse $1/L$ plot, the fit of these data to Eq.  \ref{eq:fraction}, represented as the deviation from Eq.  \ref{eq:unidist}. The fitted values, presented in Table \ref{table:energy}, were compared to the PMF values of water to provide additional validation.  By integrating the PMF function through the first layer and dividing by the width of interfacial region in the model ($\delta$), we obtained the heights and depths of the equivalent rectangular potential barriers and wells. The resulting values are also presented in Table \ref{table:energy}.  As it can be seen, in agreement with our previous findings and results reported earlier \cite{Feng2014vrd}, positive surface charge induces preferential water adsorption, while the negative surface charge causes depletion.  These data for charged surfaces neatly fits the model for the positive case, and roughly does the same for the negatively charged surface. Here the deviations under weak confinement can be partly attributed to the inherent statistical difficulties in accurately measuring small fractions with a limited number of samples.  Nevertheless, with a 5\% concentration these results should be representative as an upper limit of cases of water contamination in a charged electrochemical cell.

In the neutral case, however, data seem to follow a different route, with water levels dropping upon confinement.  This is not surprising, as from the discussion above it follows that under high confinement water molecules accumulate in the second layer (see Figure \ref{fig:figure_density}A).  This effect appears below $L=4$~nm; thus, systems with the distance between the graphene sheets over that size follow the model predictions that imply slight depletion of water near the neutral surfaces.


\section{Discussion}

The reported behaviour of water molecules in 1-butyl-3-methylimidazolium tetrafluoroborate under nanoconfinement sheds light on the behaviour of water molecules in IL-based electrolytes confined in porous carbon materials. Moreover, the proposed two-level statistical model provides an answer to the question ``how rich-in-water the electrical double layer might become, even at negligible concentrations of water in the bulk?'' raised in Ref. \cite{Fedorov2014uyy}.  The answer to this question is of great importance for engineering of supercapacitors \cite{Beguin2014tsc}, electromechanical actuators \cite{Ding2003uxv}, motion and electrochemical sensors \cite{Arrigan2015upr}, as well as energy harvesters \cite{Must2013wzs}.  In particular, our model can be applied to accelerate the transition from sealed prototypes to in-air-operable devices.  It should be noted that complete sealing, preventing the sorption of water from the air, is difficult to achieve, and this somehow inhibits the development of flexible and miniature supercapacitors or actuators. The latter were found to be especially sensitive to the humidity of the air \cite{Must2014toj}.  Wherein, the high water content in these devices has certain advantages, such as decreased viscosity of the electrolyte and, thus, increased response speed as well as bending curvatures.  Furthermore, the sensitivity to humidity turns the electromechanical actuators into good mechanical sensors, and the reversible sorption of water can be employed in energy harvesting \cite{Must2013wzs}.

Therefore, the presence of water might be beneficial, because of modification of the interfacial properties \cite{Must2014toj}, but also undesirable, due to irreversible (electro)chemical reactions \cite{Omahony2008tmf}. The presented MD simulations results demonstrate that by adjusting the confinement and surface charge density (or equivalently the potential of an electrochemical cell) one can control the presence of water in the modelled slit nanopore (or similarly in a porous carbon electrode). Noteworthy, our model considers mostly physical interactions; thus, strictly, it can be compared to a realistic system only if the water induced reactions could be neglected. Nevertheless, when fitted to the two-level statistical model the simulations results provide a clear picture of IL--water mixture behaviour under confinement, summarised below. Future studies will reveal a more detailed relationship between the electrolyte chemical structure, electrode porosity and humidity level on the ability to reversibly adsorb/desorb water and, thereby, produce electrochemical/mechanical work.


\section{Conclusions}

In the present paper we report MD simulations of 1-butyl-3-methylimidazolium tetrafluoroborate ionic liquid mixtures with water confined between two graphene sheets, \textit{i.e.} in the simplest model of a slit nanopore.  We demonstrate that water molecules are adsorbed at the surface, irrespective of the surface charge density and the height of the confinement.  The results suggest that:

\begin{itemize}
\item Due to a preferential hydration of tetrafluoroborate anions, the potential of mean force for water molecules near the positively charged surface is lower than the one for the neutral and negatively charged surfaces;
\item The translational states of water are blue-shifted at the positively charged surface due to a stronger caging in the interfacial regions than in the bulk;
\item Nanoconfinement increases the probability of the presence of water molecules near the positively charged surface while leading to depletion at the neutral and, more markedly, at the negatively charged surfaces.
\end{itemize}

The proposed two-level statistical model describes the preferential adsorption as well as a depletion of water molecules in these dense ionic environments.  This model can be used to predict the degree of water molecules accumulation at the surfaces of porous carbon materials, and, therefore, to optimise the performance of in-air-operable ionic liquid-based electrochemical and electromechanical devices.


\ack
We acknowledge the supercomputing support from the EPSRC funded ARCHIE-WeSt High-Performance Computer centre (www.archie-west.ac.uk, EPSRC grant no.  EP/K000586/1) and the Galician Supercomputing Centre (CESGA).  The financial support of the Estonian Materials Technology Program Project SLOKT12180T, Project of European Structure Funds SLOKT12026T, Estonian Institutional Research Project IUT20-013, Estonian Personal Research Project PUT1107, and Estonian Centres of Excellence in Science Project: High-technology Materials for Sustainable Development TK117 is highly appreciated.  The financial support of the Spanish Ministry of Economy and Competitiveness MAT2014-57943-C3-1-P and MAT2014-57943-C3-3-P is gratefully acknowledged.  Moreover, this work was funded by the Spanish Ministry of Economy and Competitiveness (FIS2012-33126) and by the Xunta de Galicia (AGRUP2015/11).  All these research projects were partially supported by FEDER.  Funding from the European Union (COST Action CM 1206) and by the Galician Network on Ionic Liquids, REGALIs (CN 2014/015) is also acknowledged.


\section*{References}
\bibliographystyle{iopart-num-mod}
\bibliography{WuC_article}

\providecommand{\newblock}{}
\begin{thebibliography}{10}
\expandafter\ifx\csname url\endcsname\relax
  \def\url#1{{\tt #1}}\fi
\expandafter\ifx\csname urlprefix\endcsname\relax\def\urlprefix{URL }\fi
\providecommand{\eprint}[2][]{\url{#2}}

\bibitem{Vatamanu2013ude}
Vatamanu J, Hu Z, Bedrov D, Perez C and Gogotsi Y 2013 {\em J. Phys. Chem.
  Lett.\/}  2829--2837

\bibitem{Fedorov2014uyy}
Fedorov M~V and Kornyshev A~A 2014 {\em Chem. Rev.\/} {\bf 114} 2978--3036

\bibitem{Macfarlane2014ugy}
{MacFarlane} D~R, Tachikawa N, Forsyth M, Pringle J~M, Howlett P~C, Elliott
  G~D, Davis J~H, Watanabe M, Simon P and Angell C~A 2014 {\em Energy Environ.
  Sci.\/} {\bf 7} 232--250

\bibitem{Wasserscheid2003wep}
Wasserscheid P and Welton T (eds) 2003 {\em Ionic Liquids in Synthesis\/}
  (Wiley-{VCH})

\bibitem{Ohno2005ujg}
Ohno H (ed) 2005 {\em Electrochemical Aspects of Ionic Liquids\/} (Wiley \&
  Sons)

\bibitem{Plechkova2008tnl}
Plechkova N~V and Seddon K~R 2008 {\em Chem. Soc. Rev.\/} {\bf 37}
  123{\textendash}150

\bibitem{Macfarlane2007tux}
{MacFarlane} D~R, Forsyth M, Howlett P~C, Pringle J~M, Sun J, Annat G, Neil W
  and Izgorodina E~I 2007 {\em Acc. Chem. Res.\/} {\bf 40} 1165--1173

\bibitem{Zhai2011vkf}
Zhai Y, Dou Y, Zhao D, Fulvio P~F, Mayes R~T and Dai S 2011 {\em Adv. Mater.\/}
  {\bf 23} 4828{\textendash}4850

\bibitem{Zhang2009uay}
Zhang L~L and Zhao X~S 2009 {\em Chem. Soc. Rev.\/} {\bf 38} 2520--2531

\bibitem{Simon2013ucv}
Simon P and Gogotsi Y 2013 {\em Acc. Chem. Res.\/} {\bf 46} 1094--1103

\bibitem{Beguin2014tsc}
B{\'e}guin F, Presser V, Balducci A and Frackowiak E 2014 {\em Adv. Mater.\/}
  {\bf 26} 2219--2251

\bibitem{Singh2014tpt}
Singh M~P, Singh R~K and Chandra S 2014 {\em Prog. Mat. Sci.\/} {\bf 64} 73 --
  120

\bibitem{Merlet2012wne}
Merlet C, Rotenberg B, Madden P~A, Taberna P~L, Simon P, Gogotsi Y and Salanne
  M 2012 {\em Nat. Mater.\/} {\bf 11} 306---310

\bibitem{Perkin2012tov}
Perkin S 2012 {\em Phys. Chem. Chem. Phys.\/} {\bf 14} 5052--5062

\bibitem{Rajput2014vnw}
Rajput N~N, Monk J and Hung F~R 2014 {\em J. Phys. Chem. C\/} {\bf 118}
  1540--1553

\bibitem{Ivanistsev2014tqg}
Ivani{\v s}t{\v s}ev V, {O{\textquoteright}Connor} S and Fedorov M~V 2014 {\em
  Electrochem. Commun.\/} {\bf 48} 61--64

\bibitem{Feng2013vwj}
Feng G, Li S, Presser V and Cummings P~T 2013 {\em J. Phys. Chem. Lett.\/} {\bf
  4} 3367--3376

\bibitem{Kurig2016wts}
Kurig H, Russina M, Tallo I, Siebenb{\"u}rger M, Romann T and Lust E 2016 {\em
  Carbon\/} {\bf 100} 617--624

\bibitem{Britton2014twk}
Britton J, Cousens N~E~A, Coles S~W, van Engers C~D, Babenko V, Murdock A~T,
  Ko{\'o}s A, Perkin S and Grobert N 2014 {\em Langmuir\/} {\bf 30}
  11485--11492

\bibitem{Mendez-morales2014uev}
M{\'e}ndez-Morales T, Carrete J, P{\'e}rez-Rodr{\'i}guez M, Cabeza {\'O},
  Gallego L~J, Lynden-Bell R~M and Varela L~M 2014 {\em Phys. Chem. Chem.
  Phys.\/} {\bf 16} 13271--13278

\bibitem{Ivanistsev2016vrv}
Ivanistsev V, Mendez-Morales T, Lynden-Bell R~M, Cabeza O, Gallego L~J, Varela
  L~M and Fedorov M~V 2016 {\em Phys. Chem. Chem. Phys.\/} {\bf 18} 1302--1310

\bibitem{Chen2013tnc}
Chen Y, Cao Y, Lu X, Zhao C, Yan C and Mu T 2013 {\em New J. Chem.\/} {\bf 37}
  1959--1967

\bibitem{Cuadrado-prado2009wqq}
Cuadrado-Prado S, Dom{\'i}nguez-P{\'e}rez M, Rilo E, Garc{\'i}a-Garabal S,
  Segade L, Franjo C and Cabeza O 2009 {\em Fluid Phase Equil.\/} {\bf 278}
  36--40

\bibitem{Widegren2005wre}
Widegren J~A, Laesecke A and Magee J~W 2005 {\em Chem. Commun.\/}  1610--1612

\bibitem{Rilo2009ufw}
Rilo E, Pico J, Garc{\'i}a-Garabal S, Varela L and Cabeza O 2009 {\em Fluid
  Phase Equil.\/} {\bf 285} 83--89

\bibitem{Rilo2013tvp}
Rilo E, Vila J, Garc{\'i}a-Garabal S, Varela L~M and Cabeza O 2013 {\em J.
  Phys. Chem. B\/} {\bf 117} 1411--1418

\bibitem{Omahony2008tmf}
{O{\textquoteright}Mahony} A~M, Silvester D~S, Aldous L, Hardacre C and Compton
  R~G 2008 {\em J. Chem. Eng. Data\/} {\bf 53} 2884--2891

\bibitem{Fayer2010tsk}
Fayer M~D and Levinger N~E 2010 {\em Annu. Rev. Anal. Chem.\/} {\bf 3}
  89{\textendash}107

\bibitem{Zhou2012vmt}
Zhou H, Ganesh P, Presser V, Wander M~C~F, Fenter P, Kent P~R~C, Jiang D~E,
  Chialvo A~A, {McDonough} J, Shuford K~L and Gogotsi Y 2012 {\em Phys. Rev.
  B\/} {\bf 85} 035406

\bibitem{Espinosa-marzal2014vot}
Espinosa-Marzal R~M, Arcifa A, Rossi A and Spencer N~D 2014 {\em J. Phys. Chem.
  C\/} {\bf 118} 6491--6503

\bibitem{Feng2014vrd}
Feng G, Jiang X, Qiao R and Kornyshev A~A 2014 {\em {ACS} Nano\/} {\bf 8}
  11685--11694

\bibitem{Hess2008vel}
Hess B, Kutzner C, van~der Spoel D and Lindahl E 2008 {\em J. Chem. Theory
  Comput.\/} {\bf 4} 435{\textendash}447

\bibitem{Pronk2013vhm}
Pronk S, P{\'a}ll S, Schulz R, Larsson P, Bjelkmar P, Apostolov R, Shirts M~R,
  Smith J~C, Kasson P~M, Spoel D~v~d, Hess B and Lindahl E 2013 {\em
  Bioinformatics\/} {\bf 29} 845--854

\bibitem{Van_der_spoel2010wfn}
van~der Spoel D, Lindahl E, Hess B, van Buuren A~R, Apol E, Meulenhoff P~J,
  Tieleman D~P, Sijbers A~L~T~M, Feenstra K~A, van Drunen R and Berendsen H~J~C
  2010 {\em Gromacs User Manual version 4.5\/} (www.gromacs.org)

\bibitem{Lindahl2001tps}
Lindahl E, Hess B and Van Der~Spoel D 2001 {\em J. Mol. Model.\/} {\bf 7}
  306--317

\bibitem{Berendsen1995try}
Berendsen H~J~C, van~der Spoel D and van Drunen R 1995 {\em Comput. Phys.
  Comm.\/} {\bf 91} 43{\textendash}56

\bibitem{Bussi2007vdd}
Bussi G, Donadio D and Parrinello M 2007 {\em J. Chem. Phys.\/} {\bf 126}
  014101--7

\bibitem{Spoel2005wyt}
Spoel D~v~d, Lindahl E, Hess B, Buuren A~R~v, Apol E, Meulenhoff P~J, Tieleman
  D~P, Sijbers A~L~T~M, Feenstra K~A, Drunen R~v and Berendsen H~J~C 2005 {\em
  Gromacs Users' Manual version 3.3\/}

\bibitem{Mahoney2000tfp}
Mahoney M~W and Jorgensen W~L 2000 {\em J. Chem. Phys.\/} {\bf 112}
  8910{\textendash}8922

\bibitem{Humphrey1996tsu}
Humphrey W, Dalke A and Schulten K 1996 {\em J. Mol. Graph\/} {\bf 14} 33--38,
  27--28

\bibitem{Martinez2009uoq}
Mart{\'i}nez L, Andrade R, Birgin E~G and Mart{\'i}nez J~M 2009 {\em J. Comput.
  Chem.\/} {\bf 30} 2157--2164

\bibitem{Frenkel2002wwr}
Frenkel D and Smit B 2002 {\em Understanding molecular simulation\/} (Academic
  Press)

\bibitem{Michaud-agrawal2011wef}
Michaud-Agrawal N, Denning E~J, Woolf T~B and Beckstein O 2011 {\em J. Comput.
  Chem.\/} {\bf 32} 2319--2327

\bibitem{Li2014ulb}
Li H, Wood R~J, Endres F and Atkin R 2014 {\em J. Phys.: Condens. Matter\/}
  {\bf 26} 284115

\bibitem{Ivanistsev2014wyn}
Ivani{\v s}t{\v s}ev V, Fedorov M~V and Lynden-Bell R~M 2014 {\em J. Phys.
  Chem. C\/} {\bf 118} 5841--5847

\bibitem{Ye2012woj}
Ye H, Zhang H, Zhang Z and Zheng Y 2012 {\em J. Adhes. Sci. Technol.\/} {\bf
  26} 1897--1908

\bibitem{Rotenberg2015wqi}
Rotenberg B and Salanne M 2015 {\em J. Phys. Chem. Lett.\/}  4978--4985

\bibitem{Bozym2015uyq}
Bozym D~J, Uralcan B, Limmer D~T, Pope M~A, Szamreta N~J, Debenedetti P~G and
  Aksay I~A 2015 {\em J. Phys. Chem. Lett.\/} {\bf 6} 2644--2648

\bibitem{Ding2003uxv}
Ding J, Zhou D, Spinks G, Wallace G, Forsyth S, Forsyth M and {MacFarlane} D
  2003 {\em Chem. Mater.\/} {\bf 15} 2392--2398

\bibitem{Arrigan2015upr}
Silvester D~S and Aldous L 2015 {\em {RSC} Detection Science\/} ed Arrigan
  D~W~M (Cambridge: Royal Society of Chemistry) pp 341--386 ISBN
  978-1-84973-831-6

\bibitem{Must2013wzs}
Must I, Johanson U, Kaasik F, P{\~o}ldsalu I, Punning A and Aabloo A 2013 {\em
  Phys. Chem. Chem. Phys.\/} {\bf 15} 9605

\bibitem{Must2014toj}
Must I, Vunder V, Kaasik F, P{\~o}ldsalu I, Johanson U, Punning A and Aabloo A
  2014 {\em Sens. Actuator B-Chem.\/} {\bf 202} 114--122

\end{thebibliography}

\end{document}